\documentclass[journal,twocolumn,letterpaper]{IEEE}
\ifCLASSINFOpdf
\else
\fi

\usepackage{times,epsfig}
\usepackage{fancyhdr}
\usepackage{amsmath}
\usepackage{amsfonts}
\usepackage{amssymb}
\usepackage[latin1]{inputenc}
\usepackage{lmodern}
\usepackage{array}
\usepackage{graphicx}
\usepackage{url}
\usepackage{subfigure}
\usepackage{bm}
\usepackage{breqn}
\usepackage{xcolor}
\usepackage{soul}
\usepackage{amssymb}
\usepackage{cite}
\usepackage{float}
\usepackage{multirow}
\usepackage{siunitx}
\usepackage{comment}
\usepackage{float}
\usepackage{gensymb}

\begin{document}

\title{Efficient Communication and Powering for Smart Contact Lens with Resonant Magneto-Quasistatic Coupling}

\author{Sukriti Shaw,~\IEEEmembership{Student~Member,~IEEE,}
        Mayukh Nath, 
        Arunashish Datta,~\IEEEmembership{Student~Member,~IEEE,}\\
        Shreyas Sen,~\IEEEmembership{Senior~Member,~IEEE}       
}

\twocolumn[
\begin{@twocolumnfalse}
  
\maketitle

\begin{abstract}
A two-coil wearable system is proposed for wireless communication and powering between a transmitter coil in a necklace and a receiver coil in a smart contact lens, where the necklace is invisible in contrast to coils embedded in wearables like spectacles or headbands. Magneto-quasistatic(MQS) field coupling facilitates communication between the transmitter in the necklace and the contact lens receiver, enabling AR/VR and health monitoring. As long as the receiver coil remains within the magnetic field generated by the transmitter, continuous communication is sustained through MQS field coupling despite the misalignments present. Resonant frequency tuning enhances system efficiency. The system's performance was tested for coil misalignments, showing a maximum path loss variation within $10 dB$ across scenarios, indicating robustness. Finite Element Method(FEM) analysis has been used to study the system for efficient wireless data transfer and powering. A communication channel capacity is $4.5 Mbps$ over a $1 MHz$ bandwidth. Simulations show negligible path loss differences with or without human tissues, as magnetic coupling remains unaffected at MQS frequencies below $30 MHz$ due to similar magnetic permeability of tissues and air. Therefore, the possibility of efficient communication and powering of smart contact lenses through a necklace is shown for the first time using resonant MQS coupling at an axial distance of $15cm$ and lateral distance of over $9cm$ to enable AR/VR and health monitoring on the contact lens.

\end{abstract}


\end{@twocolumnfalse}]

{
  \renewcommand{\thefootnote}{}%
  \footnotetext[1]{This work was supported by the National Science Foundation Career Award No. 1944602.}
  \footnotetext[2]{S. Shaw, A. Datta and S. Sen are with the Department of Electrical and Computer Engineering, Purdue University, West Lafayette, IN 47907 USA (email: shaw161@purdue.edu, datta30@purdue.edu, shreyas@purdue.edu).}
  \footnotetext[3]{M. Nath was with Purdue University, West Lafayette, IN 47907 USA during the time of this work (email: nathm@alumni.purdue.edu).}
}

\IEEEpeerreviewmaketitle

\section{Introduction}
\IEEEPARstart{T}{he} growth of smart devices has significantly enhanced the convenience of modern life with individuals equipped with wearables and implants on themselves, be it for physiological signal monitoring or for getting instant access to information about personal matters or global events on their fingertips. However, this accessibility has also led to a shift in human interaction encroaching on social gatherings with individuals increasingly relying on virtual connections rather than engaging in face-to-face communication thereby diminishing the quality of interpersonal interactions. The integration of interconnected devices and AI assistants has extended beyond stationary settings to include mobile environments, requiring seamless connectivity on the go.

\begin{figure}[t]
    \centering
    \includegraphics [width=1\linewidth] {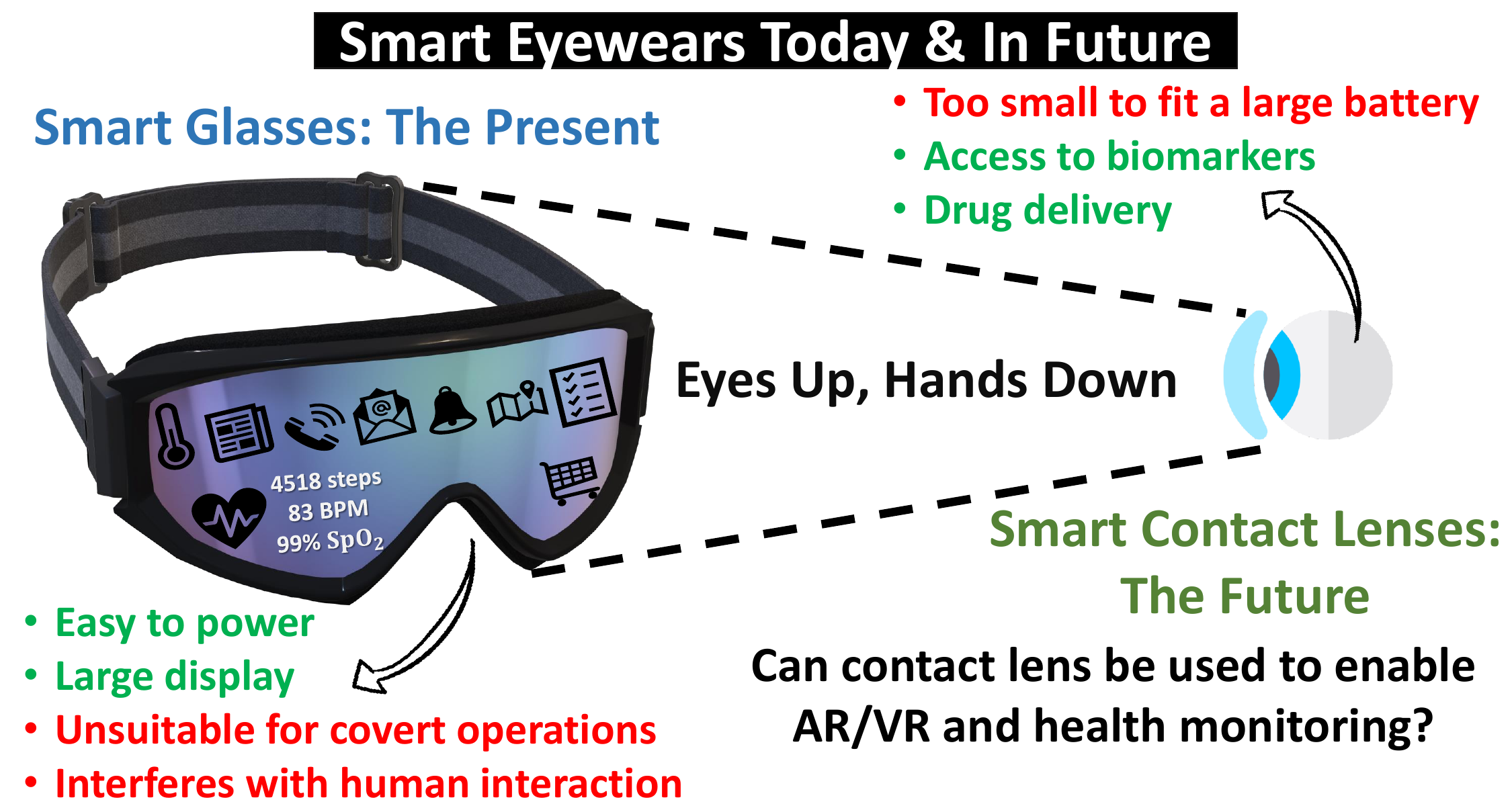}
    \caption{Goal towards miniaturizing smart eyewear}
    \label{fig_intro}
    \vspace{-0.15in}
\end{figure}

To promote an "eyes up, hands down" lifestyle, smart eyewear such as glasses and contact lenses offer a promising solution. These innovative wearables can display a range of information directly in the user's field of vision, eliminating the need to constantly check a smartphone for updates such as grocery lists, news headlines, or medical reminders as listed in Figure \ref{fig_intro}. While the development of smart glasses, \cite{metasmartglasses},\cite{raybaniqbal2023adopting}, is advancing rapidly, challenges remain. These devices require frequent charging to sustain their displays, and their bulkiness may pose discomfort to users, particularly during busy daily routines. However, as technology continues to evolve, smart glasses are poised to become more commonplace, offering enhanced functionality and convenience.

With the multitude of these devices on the body communicating with each other, not only must they have low power and low channel loss with a high data rate but they must also be hassle-free, less bulky, and comfortable for the users. Various wireless power transfer(WPT) techniques have been used to charge devices without the need to remove the devices from their usage position, especially in cases of perpetually operating devices such as implantables. \cite{yoo2021wireless} explores the various methods of WPT for properly functioning biomedical devices. 

Smart contact lenses exemplify miniature wearables and implantables, necessitating streamlined circuitry to address the challenge of bulkiness. Designing smart contacts faces a bottleneck in accommodating circuitry for heavy computation within limited dimensions. \cite{surveyxia2022state} illustrate the first innovation to make contact lenses more than a vision correction wearable back in 2003. Since then the idea to make smart contact lenses practical has come forward with techniques for enabling communication and powering the lenses. \cite{surveyshaker2023future} discusses several communication and powering techniques for contact lenses. Various energy harvesting methods, including external, in vivo, and hybrid techniques, have also been discussed in \cite{mirzajani2022powering}. While these techniques offer reliable solutions for continuous powering, their implementation remains challenging. Despite integrating wireless circuits, glucose sensors, and displays in a contact lens, as demonstrated by \cite{park2018soft}, communication and powering still necessitate proximity to an RF energy source, thus lacking perpetual operation. 
\cite{rf_coils} analyses the lateral and angular misalignment effects in RF coil systems for implantables for coil dimensions of comparable dimensions.   
\begin{figure}[t]
    \centering
    \includegraphics [width=1\linewidth] {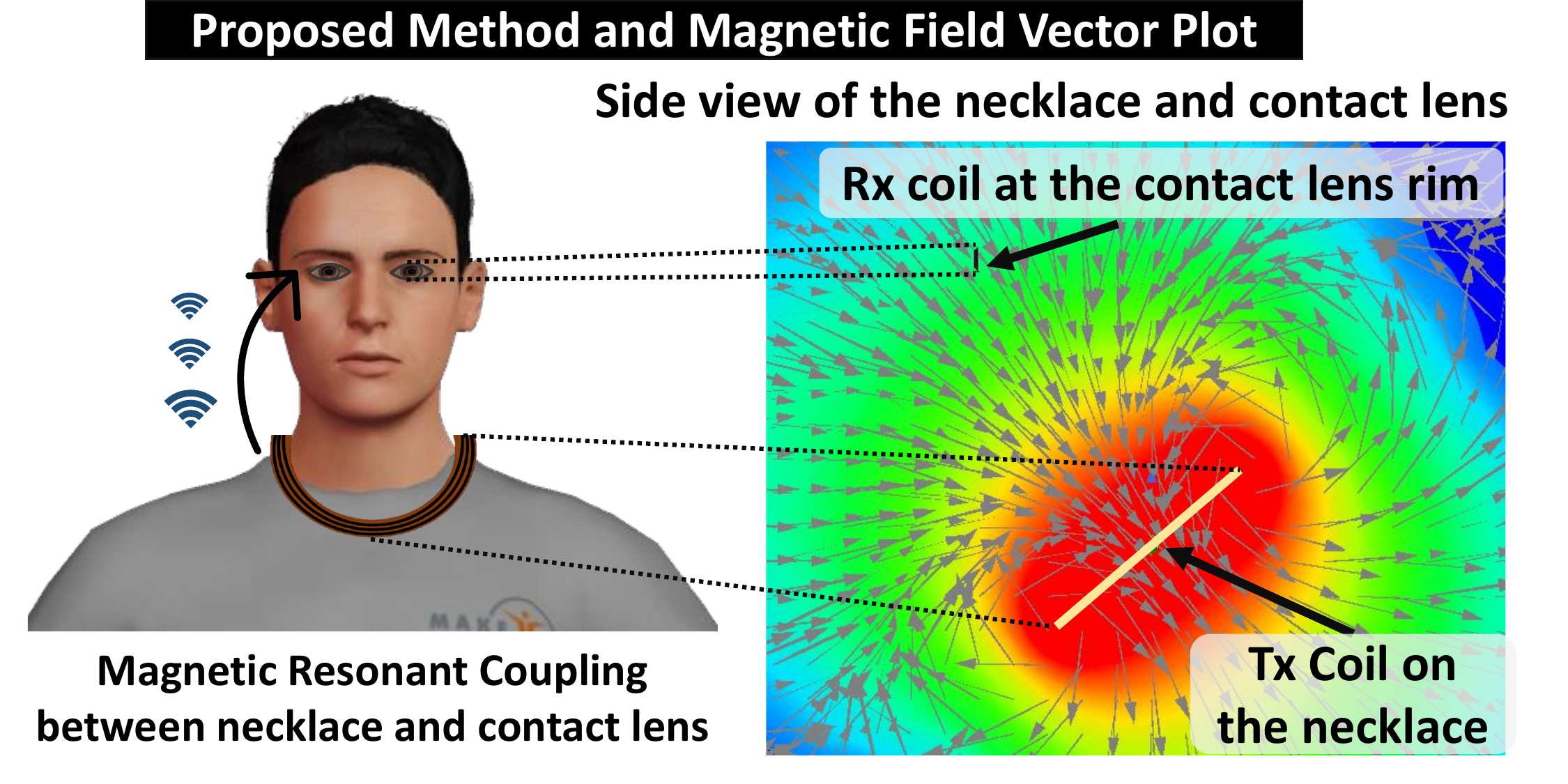}
    \caption{Proposed system based on Magnetic resonant coupling}
    \label{proposed}
    \vspace{-0.3in}
\end{figure}

WPT, as illustrated in \cite{zhang2018wireless}, \cite{fotopoulou2010wireless}, \cite{adeeb2012inductive}, and \cite{degen2021inductive}, utilizes inductively coupled links as a prevalent technique for powering and charging wearable and smart devices. In these scenarios, flat spiral coils serve as the inductors, facilitating power transfer through electromagnetic interaction between transmitter and receiver coils. The configuration, positioning, and dimensions of these coils are crucial for enabling power transfer and communication within coil systems and vary depending on applications related to health monitoring. Communication and powering of smart contact lenses through inductive coupling for the applications as illustrated in Figure \ref{fig_intro} have been explored by \cite{mobilechen2015energy}, \cite{chen2017warpage}, \cite{takamatsu2019highly}, and \cite{zada2021simultaneous} where the distance between the transmitter and the receiver are a few tens of millimeter with the coils axially aligned. The power transmission efficiency(PTE) achieved depends on the size of the coils and the distance between them. These studies on smart contacts for health monitoring applications based on biomarkers, eye pressure, and drug delivery have different requirements of power \cite{ma2021smart}, \cite{yuan2021wireless} when compared to smart contacts used for video processing and enabling Augmented Reality. These bring out the need for higher power and efficient high-speed communication. 

This work studies the resonant MQS coupling system for smart contact lenses, as patented in \cite{sen2023communication} utilizing a necklace comprising a transmitter coil and a receiver coil positioned on the rim of the contact lens as shown in Figure \ref{proposed}. The goal of this paper is to develop the theory, supported by 3D Finite Element Method(FEM) simulations using the High-Frequency Structure Simulator(HFSS) to lead to possible maximum communication efficiency and data rate, wireless power transfer efficiency in the MQS region, i.e., below $30MHz$, from the necklace to the contact lens as a function of coil dimensions, lateral and angular alignments, and separation distances. A proof of concept is presented based on the fundamental parameters: coil sizing, placement, and orientation, and an optimized model is identified to efficiently transmit data from the necklace to the lens and power the contact lens. To the best of our knowledge, this work is the first of its kind to demonstrate efficient communication based on MQS coupling for heavily misaligned, greater than $10cm$ separation distanced coil systems with dissimilar dimensions to enable AR/VR in smart contact lenses. 

\section{Theory}

\subsection{Electromagnetic Induction} 
Any alternate current-carrying conductor can induce an electromotive force in another conductor placed in the region of the magnetic field lines the former produces. The magnetic flux created at the receiver reduces as the receiver distance from the source increases. A refined formula for calculating the near-zone magnetic field of small circular transmitting loop antennas was developed by \cite{greene1967near}, incorporating corrections for frequency, finite propagation time, and loop radii. \cite{simpson2001simple} derives the magnetic field for a circular current loop which is exact throughout all space outside the conductor. The limit of the magnetic flux that induces the EMF depends on factors such as the size of the conductors, the distance of separation between the two conductors, and the alignment of the coils. The following sections shall define expressions for multi-turn coils and how the proposed system is optimized based on the same. 

The direction of the magnetic field lines produced due to a current-carrying circular coil is perpendicular to the plane of the coil and the field lines extend at infinity at the center of the coil while it bends to form a loop near the conducting wires. The magnetic field lines are shown in Figure \ref{proposed} in the form of the vector and field plot for a current-carrying circular flat spiral coil with multiple turns and the transmitter aligned at an angle while the perpendicular receiver moved away from the center of the coil. 

\subsection{Factors affecting MQS Coupling}

\subsubsection{Coil Dimensions} The number of turns, coil radius, wire diameter, and wire spacing change the coupling coefficient between the transmitter and the receiver. Similar-sized coils have a higher coupling efficiency as compared to coil systems with highly varied diameters. As the number of turns of the coils are increased the coupling between the coils increases but increasing it further leads to reduced efficiency due to the increase in parasitic capacitances between the wires. As the radius of the coil reduces, the concentration of magnetic field lines reduces leading to lower coupling. The variation in the radius of the coil and number of turns changes the inductance, series resistance, and parasitic capacitance of the inductor which affects the overall impedance and the quality factor of the circuit.

\subsubsection{Alignment} For an inductively coupled system, the alignment of the coils plays an important role in the efficiency of the system. The maximum efficiency can be achieved when the coils are similarly sized and axially aligned such that the maximum field gets coupled to the receiver. In the proposed system, there are restrictions based on the sizing of the coils, lateral shift, separation distance, and the orientation in which the coils are placed. The transmitter coil which is supposed to be attached like a necklace will be aligned at an angle. In contrast, the receiver coil is placed as far as a necklace would be from the eyes aligned at a fixed angle with the inner diameter dimensions such that the visibility from the contact lens is not hampered. The robustness of the system is verified in the later sections taking into consideration, the misalignments and prone to movement scenarios.

\subsubsection{Source and Load Resistances}
As the source resistance, $R_{Source}$, is reduced, more current flows through the transmitter coil resulting in higher power consumption at the transmitter but more power is transferred to the receiver coil. The magnetic field strength is directly proportional to the current flow. Hence the increase in the magnetic flux through the transmitter coil leads to increased induced EMF producing higher current at the receiver resulting in higher received voltage. The increase in the load resistance, $R_{Load}$, increases the voltage picked up by the receiver coil.

\subsection{Coil Design Equations}
The factors that affect the coupling coefficient come from the mutual inductance between the coils, which is dependent on the self-inductance of the coil \cite{waters2014optimal}. For flat spiral coils, the inductance $L$ can be calculated from the modified Wheeler's expression, \cite{hussain2021self} with the number of turns, $N$, the outer diameter of the coil, $D_o$, the wire diameter, $d$, and the wire spacing, $s$, given as (ref{1})
\begin{equation}
    \label{1}
    L = \frac{N^2(D_o - N(d+s))}{16D_o +28N(d+s)} \frac{39.37}{10^6} 
\end{equation}
The above expression is invalid when $N$ is small, $d>>s$, and when the ratio between the radial depth and the winding radius is less than 0.2 as mentioned in \cite{waters2014optimal}. When these conditions are defied, the expression (\ref{2}), based only on the arithmetic and geometric mean distances and the arithmetic mean square distance \cite{hussain2021self} is considered to be applied to coils with large gaps in between and turns not uniformly distributed throughout the diameter of the coil.
\begin{equation}
    \label{2}
    L = \frac{\mu_0 N^2 (\frac{D_o+D_i}{2})}{2} (\ln{\frac{2.46}{\gamma}} + 0.2\gamma^2),
\end{equation}

\begin{equation}
    \label{3}
    \gamma = \frac{D_o-D_i}{D_o+D_i}
\end{equation}
where $D_i$ is the inner diameter of the coil.

\begin{figure}[b]
    \centering
    \includegraphics [width=1\linewidth] {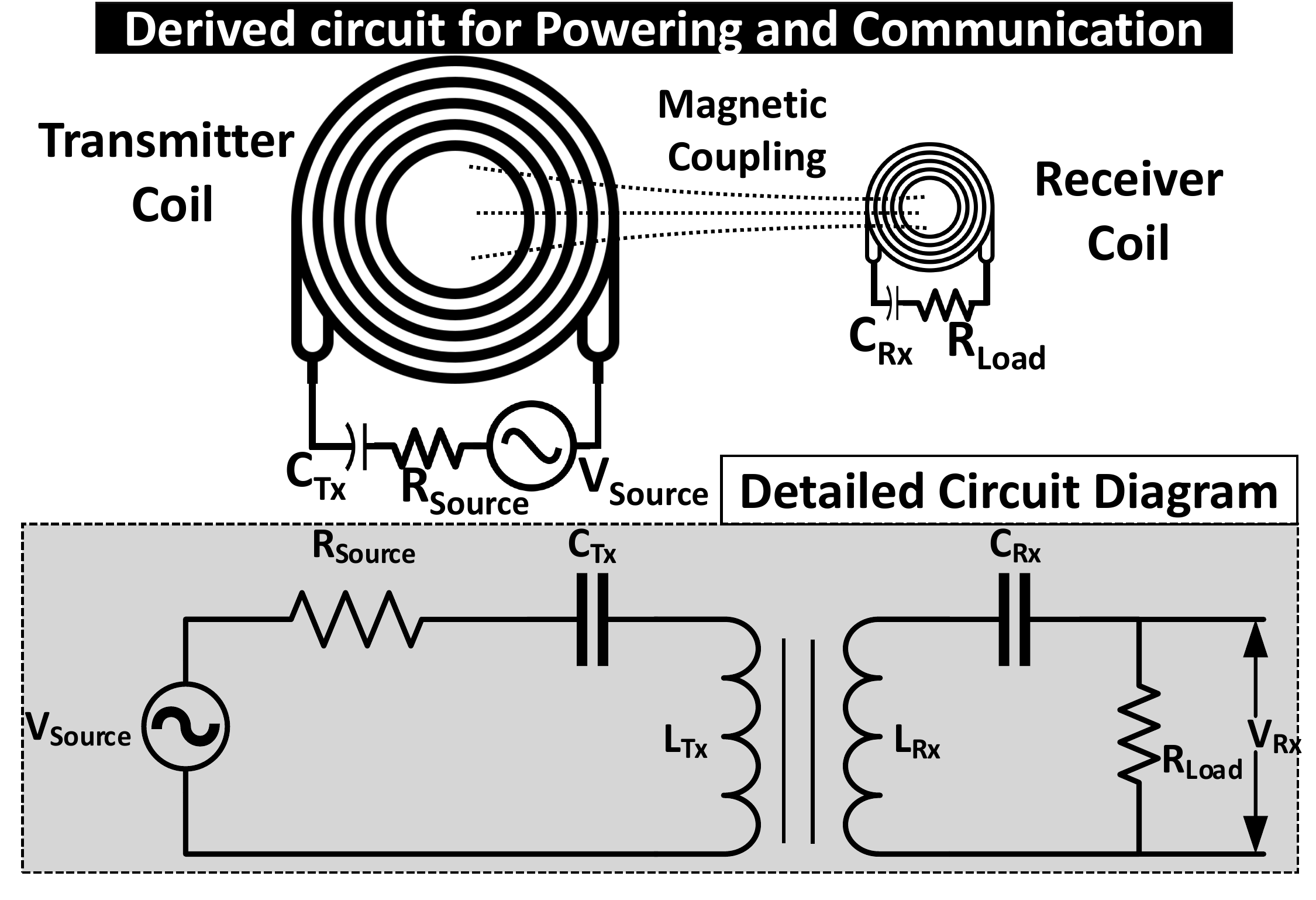}
    \caption{Circuit based on magnetic resonant coupling used to show that wireless power transfer is possible. Here the flat spiral circular coils are depicted by $L_{Tx}$ and $L_{Rx}$}
    \label{circuit}
    \vspace{-0.2in}
\end{figure}
\begin{figure*}[h]
    \centering
    \includegraphics [width=1\linewidth] {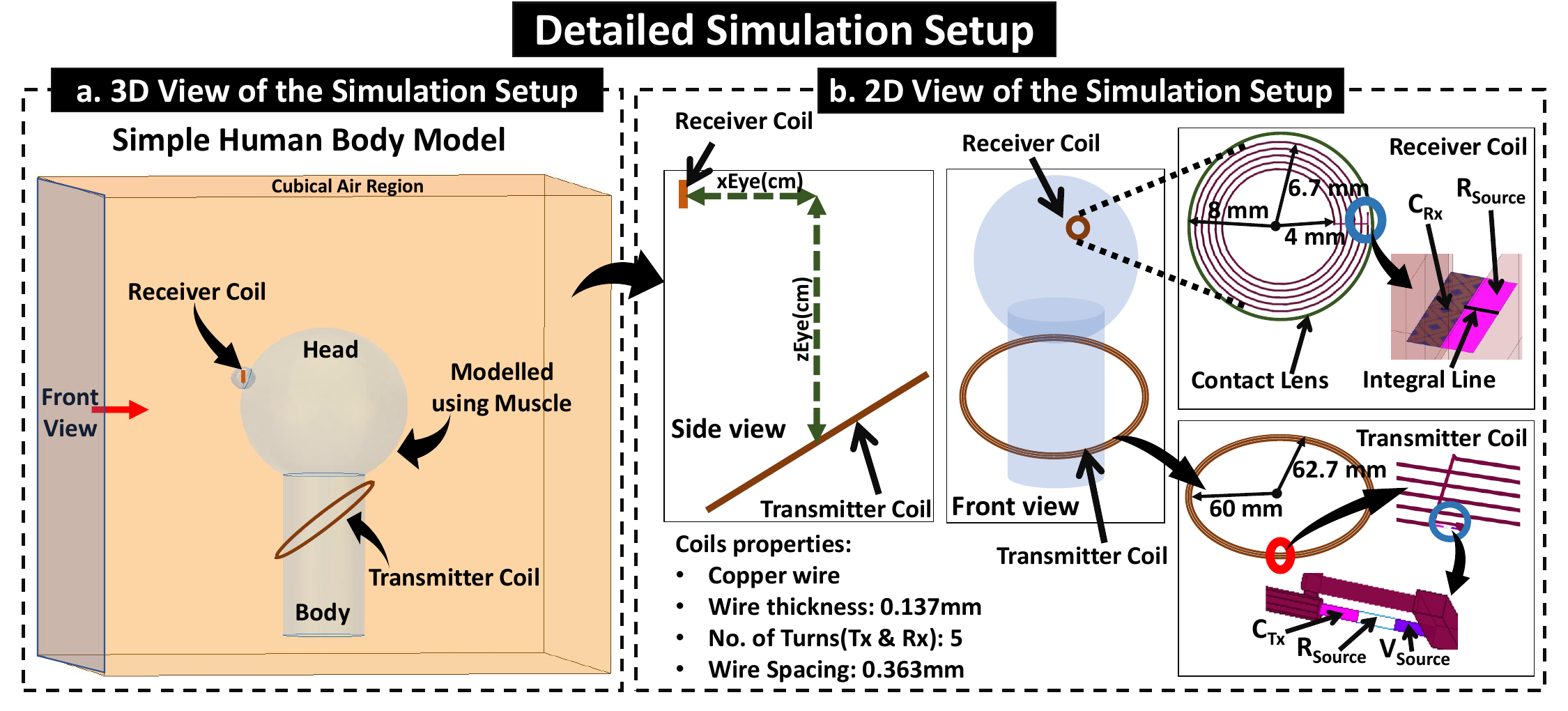}
    \caption{a. The side view shows how the coils are arranged in the simulation setup to mimic the position of a necklace and a contact lens while the front view shows the structure of the coil with the parameters for the chosen optimized model; c. 3D view of the simulation setup showing the simple human body model}
    \label{setup}
    \vspace{-0.2in}
\end{figure*}
The mutual inductance for coils arbitrarily aligned coil as described in \cite{golestani2018theoretical} is used to calculate the coupling coefficient of the proposed misaligned system to enable maximum voltage transfer. A simplified expression for the mutual inductance is given in terms of the $\phi_{Tx}$, the flux induced in the receiver coil due to the transmitter coil. 
\begin{equation}
    \label{7}
    M = N_{Rx} \frac{\phi_{Tx}}{I_{Tx}} = N_{Tx} \frac{\phi_{Rx}}{I_{Rx}} 
\end{equation}
\begin{equation}
    \label{8}
    \phi_{Tx} = \int_{S_{Rx}} B_{Tx}.\,dS_{Rx}
\end{equation}
where $S_{Rx}$ is the area of the receiver coil and $B_{Tx}=\mu H_{Tx}$ is the magnetic flux density generated by the transmitter coil. The closed-form expression for mutual inductance for coil systems as a function of the position of the coil and angular alignment has been discussed in detail by \cite{golestani2018theoretical} which considers the effect of frequency, coil geometry, misalignment, distance, and environment.

As the mismatch in coil dimensions increases, the coupling coefficient, $k$, decreases resulting in reduced induced voltage due to the reduction in the mutual inductance. The expression for the coupling coefficient is given by 
\begin{equation}
    \label{9}
    k=\frac{M}{\sqrt{L_{Tx}L_{Rx}}}
\end{equation}

The voltage transfer ratio between the transmitter and the receiver can be calculated using: 
\begin{equation}
    \label{10}
\frac{V_{Rx}}{V_{Tx}} = \frac{j\omega M R_{Load}}{(j\omega L_{Tx} + R_{Source})(j\omega L_{Rx}+ R_{Load}) + \omega^2 M^2}
\end{equation}
where $\omega$ is the angular frequency.

The resistance, $R$ of the coil will contribute to the quality factor of the coil and is dependent on the skin depth, $\delta$, and proximity effect.
\begin{equation}
    \label{4}
    R = \sqrt{\frac{f\pi\mu_o}{\sigma}} \frac{N(D_o - N(d+s))}{d}
\end{equation}
\begin{equation}
    \label{5}
    \delta = \frac{1}{\sqrt{\pi f \sigma \mu_o}}
\end{equation}

where $f$ is the operating frequency, $\mu_o$ is the permeability of free space, and $\sigma$ is the conductivity of the wire.

The quality factor $Q$ for the series resonant flat spiral coil is defined as 
\begin{equation}
    \label{6}
     Q = \frac{\omega L}{R}
\end{equation}
The quality factor of the coils increases if the coils have a larger diameter which reduces the internal resistances of the inductors. A higher quality factor leads to lower losses and dissipation.
\subsection{Resonant MQS Coupling}
Figure \ref{circuit} represents the circuit model of the proposed system with $L_{Tx}$ and $L_{Rx}$ representing the transmitter and receiver flat spiral coils, respectively. $R_{Source}$ is used to limit the current coming from the transmitting device, i.e., a 1V sinusoidal signal, $V_{Source}$. The capacitance, $C_{Tx}$, is used for tuning the resonant frequency based on the inductance, $L_{Tx}$. 
On the receiver end, $C_{Rx}$ is based on the expression mentioned below:

\begin{equation}
    \label{11}
    L_{Tx} C_{Tx} = L_{Rx} C_{Rx}
\end{equation}

$R_{Load}$ is used as the resistor across which the received voltage, $V_{Rx}$ is picked up, and hence power is measured according to 
\begin{equation}
    \label{12}
    P_{Rx} = \frac{V_{Rx}^2}{R_{Load}}
\end{equation}

Apart from the tuned resonating frequency of the coils, there will be resonant peaks due to the self-resonance of the coils acting as antennas.

\section{Methodology and Design}
A two-coil inductively coupled system is considered wherein the transmitter coil is looped around a necklace while the receiver coil is at the circumference of the contact lens to enable seamless and unobstructed experience. The coil diameters are chosen according to the necklace and contact lens dimensions. The number of turns of the coils is fixed such that maximum coupling occurs taking into effect the parasitic capacitance and the quality factor of the inductors. The detailed simulation model and setup designed in Ansys' High-Frequency Structure Simulator(HFSS) are discussed in the following subsections.

\subsection{Simulation Setup}
A cubical air region, depicted in Figure \ref{setup}a is created for the simulations to be confined to a finite space. The human body model comprising of the muscle and eye tissues is placed at the center with the transmitter coil wound around like a necklace and the receiver coil placed on the edge of the contact lenses in the eye. The thickness of the 
wires for the transmitter and the receiver coil has been considered small so that the receiver does not occupy a large area of the contact lens and the transmitter coil is lightweight. The flat spiral coils are made up of 36 American Wire Gauge(AWG) copper wire with a diameter of $0.137mm$, an inner radius of $6cm$ for the Tx and $0.4cm$ for the Rx, and a wire spacing of $0.5mm$ between each turn is shown in Figure \ref{setup}b. 

The two ends of the coil are treated as the ends of an inductor where the transmitter is excited with a sinusoidal wave of 1V and a frequency of 30MHz with a source resistance in series along with the capacitor for tuning the circuit at a particular resonant frequency. Without the capacitor, the system becomes non-resonant. The receiver coil ends are connected with a capacitor in series with a load resistor from where the voltage is picked up. The equivalent circuit diagram is shown in Figure \ref{circuit}, where the flat spiral coil is depicted in the form of inductors. 
\begin{figure}[t]
    \centering
    \includegraphics [width=1\linewidth] {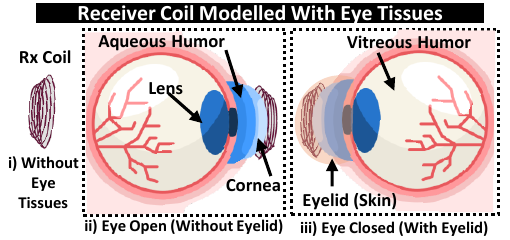}
    \caption{The eye tissues used for the simulations were cornea, aqueous humor, lens, and vitreous humor. The skin covers the receiver coil to depict the eyelid.}
    \label{eye_tissue}
    \vspace{-0.2in}
\end{figure}
\subsection{Modelling of the Eye Tissues}
Considering the eye's curvature, the receiver coil has been made helical as shown in Figure \ref{eye_tissue} to engulf the contact lens. The dimension and the radius change remain the same as \ref{setup}b. Due to the small coil dimensions, there was a negligible change in the inductance of the receiver coil which in turn does not affect the response as seen in the following sections. The tissues used to model the eye are labeled in Figure \ref{eye_tissue} \cite{zada2021simultaneous}. The receiver is placed on top of the cornea. The aqueous humor lies just behind the cornea and is considered as it is made with the most conductive body tissue, the cerebrospinal fluid(CSF) to analyze whether there is absorption of magnetic field leading to reduced channel efficiency. To mimic the eyelid for the eye-closed case, a layer of skin is placed on top of the receiver coil.

\subsection{Optimization Techniques to Maximize Coupling}
The inductance calculated using (\ref{1}) also depends on the ratio of the winding radius and the pitch radius of the coils and is applicable and correct only when the ratio is greater than 0.2 \cite{waters2014optimal}. The inductance calculation for the receiver coil using the flat spiral coil expressions (\ref{1}) and (\ref{2}) matches the measured simulated value of $0.4\mu H$ while for the transmitter coil, the inductance value is calculated using (\ref{1}) and (\ref{2}) has a huge difference when compared to the measured simulated value of $35\mu H$. To verify the calculated value with the simulation measurements, the ends of the coil have been excited with a lumped port, and the imaginary impedance parameter, $Z_{11}$ is measured over a frequency range up to $30MHz$. The imaginary $Z_{11}$ is found to be positive, that is, the reactive component is used to calculate the inductance $L$ of the coil according to the expression (\ref{13}). 
\begin{equation}
    \label{13}
    L = \frac{Z_{11}}{2 \pi f}
\end{equation}
where $f$ is the frequency at the $Z_{11}$ used.

The measured transmitter inductance does not match the calculated value since the Tx inductor has a large gap in between making the inductance equations inaccurate to calculate. Further analysis is required to measure the inductances of coils such as the transmitter coil in the proposed system which is currently beyond the scope of this work. Hence, the simulated inductance value is considered for tuning the device. On tuning the device with the measured $L_{Tx}$ and a particular resonant frequency, $\omega$ of $26 MHz$ is fixed by setting the series capacitance, $C$ at the transmitter. The external capacitance should be greater than the parasitic capacitance of the transmitter for it to make an effect on the resonant frequency \cite{golestani2018theoretical}, and determined using (\ref{14}).
\begin{equation}
    \label{14}
    C = \frac{1}{{\omega}^2 L}
\end{equation}

The position of the necklace and the contact lens are taken into consideration with the transmitter coil aligned at an angle in the way a necklace is worn around while the receiver coil is placed at the rim of the contact lens on the eye, perpendicular to the plane of the transmitter when it is not aligned at an angle. Figure \ref{setup}a depicts the configuration of the inductively coupled two-coiled system. As the necklace's position is not fixed and is prone to angular movement, the transmitter angle is varied to check for variation in the received voltage. The number of turns is fixed at 5 for both the transmitter and the receiver.  
\begin{figure}[b]
    \centering
    \includegraphics [width=1\linewidth] {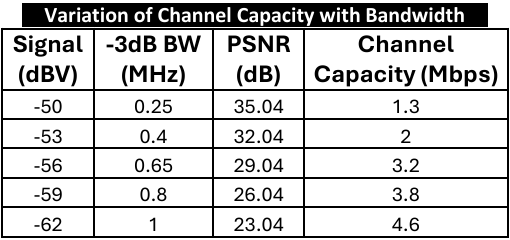}
    \caption{Calculation of channel capacity by varying the 3dB bandwidth; the noise floor is assumed to be $-85dBV$ }
    \label{SNR}
    \vspace{-0.2in}
\end{figure}
\subsection{Communication Channel Capacity}
The maximum theoretical bit rate of a wireless communication channel is given by the Shannon-Hartley theorem defined by the channel capacity. The robustness and efficiency of the communication channel formed by the resonant MQS coupling are measured in terms of the maximum bit rate it can achieve which would show possibilities enabling video transfer for AR/VR on the contact lens. The channel capacity, $C_{ch}$ is given by
\begin{equation}
    \label{15}
    C_{ch} = BW(\log_2 ({1+SNR}))
\end{equation}
where the $BW$ is the 3 dB communication bandwidth and the voltage signal-to-noise ratio is given by the $SNR$. The noise floor is assumed to be $-85dBV$, the voltage level of the non-resonant flatband signal obtained. The SNR calculation is based on the transmitting signal amplitude of $1V$. For improving the channel capacity, the transmitting signal amplitude can be increased further to increase the SNR or the $BW$ can be increased to a level such that the signal remains higher than the noise floor. 

To increase the signal's bandwidth, the signal level is further reduced to keep the PSNR within the detectable limits as seen in Figure \ref{SNR}. Reducing the signal level by $12dB$ increases the channel capacity to over $4.5Mbps$.

\section{Results}

\subsection{MQS and Resonant MQS Coupling}
\begin{figure}[b]
    \centering
    \includegraphics [width=1\linewidth] {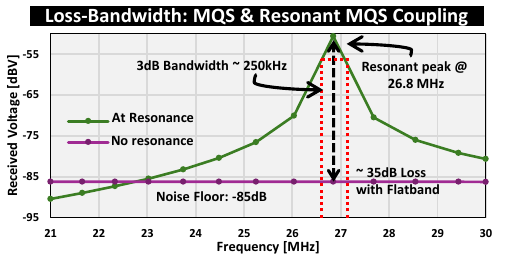}
    \caption{Frequency sweeps for the untuned and tuned transmitter circuits with the coils suspended in the air}
    \label{mqsLoss}
    \vspace{-0.2in}
\end{figure}

When the transmitter coil, $L_{Tx}$, is directly shorted to the $R_{Source}$, it is not tuned at any frequency making the coupling non-resonant. The dimensions and the positions of the transmitter and receiver coil are fixed as discussed in the section above and shown in Figure \ref{setup}b. Depending on whether the coils are connected in series with a capacitor $C_{Tx}$ and $R_{Source}$ or not, the coupling would be resonant or non-resonant. Figure \ref{mqsLoss} shows the frequency response for the two cases. It is seen that path loss for the resonant MQS coupling is reduced giving a gain of over 35dB as compared to the non-resonant coupling which has a flatband while the 3dB bandwidth of the resonant MQS coupling is almost $1MHz$. For the following sections, -85dB is considered as the noise floor for the wireless communication channel. The communication bandwidth of the channel can be further increased to 1 MHz facilitating a higher data rate. The resonant efficiency boosts the communication efficiency making the channel suitable for high-speed video transmissions. 
\begin{figure}[t]
    \centering
    \includegraphics [width=1\linewidth] {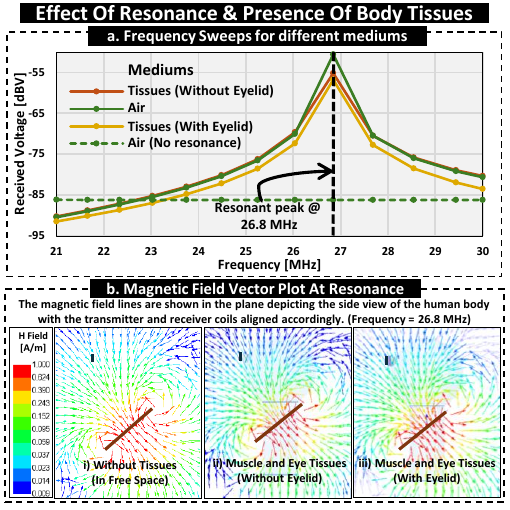}
    \caption{The receiver picks up almost the same voltage in different mediums when the circuit is resonant}
    \label{sweep}
    \vspace{-0.2in}
\end{figure} 

\subsection{Presence of human tissues}

When the transmitting device is activated, the current flowing through the transmitter coil generates a magnetic field following the right-hand thumb rule, inducing a current in the receiver coil for wireless communication and power transfer. The coils are subjected to several mediums to observe the effect of body tissues on the communication channel between the transmitter and the receiver coils. The coils are freely suspended in the air without any human body model, then the transmitter coil is wound around the head and body model made up of the muscle tissue while the receiver coil is attached to the eye tissues as shown in Figure \ref{eye_tissue}(ii). To simulate the case for closed eyes, the eyelid made up of skin is placed on top of the receiver coil, Figure \ref{eye_tissue}(iii).

Figure \ref{sweep}a shows the frequency sweeps for the non-resonant mode in the air, the resonant circuits in free space, the muscle and eye tissues without the skin, and the tissues with the eyelid. It is observed that the receiver picks up almost the same voltage in all the 3 cases of the resonant mode with minimal absorption. Figure \ref{sweep}b depicts the magnetic field patterns produced by the 3 mediums for the resonant circuit at $26.8MHz$. The similarity in field strength indicates minimal absorption of magnetic field lines by the muscle tissue at the operating frequency, attributed to its relative permeability ($\mu_{r}$) of 1, rendering it unaffected by the surrounding magnetic field lines \cite{nath2022understanding}. However, as the frequency increases, the conductivity of human body tissues rises, leading to the generation of eddy currents due to the magnetic field, resulting in the absorption of the field by the tissues and subsequent reduction in received voltage levels.

\begin{figure}[t]
    \centering
    \includegraphics [width=1\linewidth, scale=1] {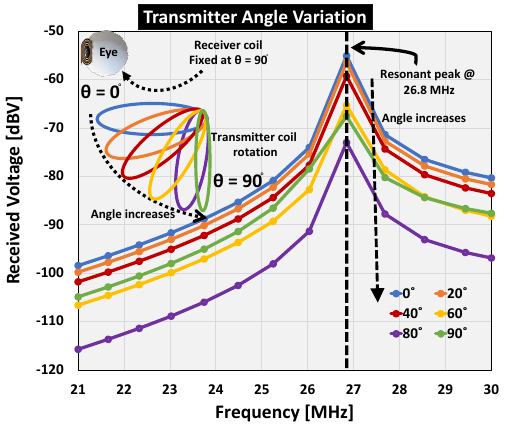}
    \caption{The various positions of the transmitter coil angle along with the received voltage frequency sweeps}
    \label{TxAngle}
    \vspace{-0.2in}
\end{figure} 
\subsection{Robustness of the Proposed System}

As mentioned in the previous sections, in the proposed modality for enabling efficient communication and powering to smart contact lenses with a resonant MQS coupled system, the coils would be highly misaligned. This work studies the possibilities for such a system and finds the maximum possible communication channel efficiency to enable AR/VR on contact lenses. The axis of the receiver coil is always considered perpendicular to the transmitter.

As the transmitter coil is wound resembling a necklace, its orientation can vary due to movement, potentially inclining at different angles rather than being fixed. Similarly, the perpendicular axial distance and the lateral shift between the transmitter and the receiver coil shall vary across individuals depending on their body structure and dimensions. Figure \ref{robust}a shows the various misalignment positions used to sweep across for checking the variation in channel capacity. $xEye=92mm$, $zEye=150mm$, and Tx Angle=$40\degree$ are the ideal positions and orientations for the coils. Some positions of the coils are unrealistic in the case of the proposed system but they can still be used to analyze the misalignment of such inductively coupled systems.

\subsubsection{Angular Misalignment}

The transmitter inclination angle ranges from  $0\degree$(parallel to the ground) to $90\degree$(perpendicular to the ground). While having the coil completely parallel or perpendicular to the ground is not optimal, angles within the range $20\degree-60\degree$ exhibit a variation of approximately $8dB$. Considering a circular necklace, an angle of $40\degree$ has been deemed optimal for all simulations to demonstrate the feasibility of wireless communication power transfer between a necklace and a contact lens. Figure \ref{TxAngle} illustrates various necklace positions alongside the resonant peaks corresponding to transmitter angle variations. It is seen that the transmitter at $80\degree$ has a higher path loss than at $90\degree$ due to the magnetic fields canceling out each other owing to the axes of the coils being perpendicular to each other.

Figure \ref{robust}b shows the variation of channel capacity at $26.8MHz$ concerning the angular misalignment of the transmitter coil. It can be concluded that even if there is a $10dB$ variation in the received signal level, the channel capacity does not go below $4.5Mbps$ with the minimum bandwidth considered.

\subsubsection{Lateral Misalignment}

For the lateral movement of the receiver, the channel capacity is measured in cases when the transmitter coil is inclined at angles $0\degree$ and $40\degree$, it is observed that the theoretical bit rate remains over $4.5 Mbps$ as shown in Figure \ref{robust}c, indicating robustness. The range chosen is between $60-140mm$ to show variations among different human dimensions for coupling among the coils.

\subsubsection{Axial Misalgnment} 
\begin{figure}[b]
    \centering
    \includegraphics [width=1\linewidth] {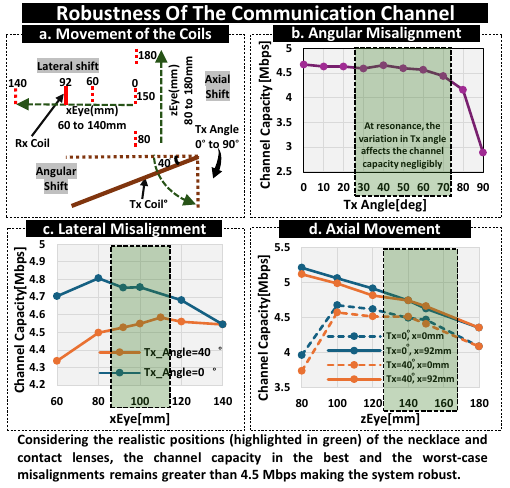}
    \caption{The variation in channel capacity with a) angular, b) lateral, and c) axial misalignment}
    \label{robust}
    \vspace{-0.1in}
\end{figure}

The coupling between the axially aligned coils increases as the distance between them reduces but when their axes are perpendicular to each other, the magnetic field lines tend to couple less strongly. In Figure \ref{robust}d, all three misalignments are taken into consideration. For $xEye=0mm$(not practical), it is seen that the field lines coupled eventually reduce as the distance between the coils increases while for $xEye=92mm$, the bit rate decreases almost linearly. It is seen that the variation in the transmitter angle is not dominated in this plot. Hence, for realistic scenarios, the channel is robust with a bit rate of over $4.5Mbps$.

\subsection{Varying Impedances for Dual Mode Operation}

The impedance on the source side typically serves to either regulate the current or indicate the presence of some impedance in the circuit that restricts the current flow. This includes the wire resistance between the transmitting device and the subsequent component, which adds to the source impedance. To gauge the received voltage, the load impedance is employed to enable the measurement of voltage drop. 
\subsubsection{Effect of Source Impedance}
\begin{figure}[t]
    \centering
    \includegraphics [width=1\linewidth] {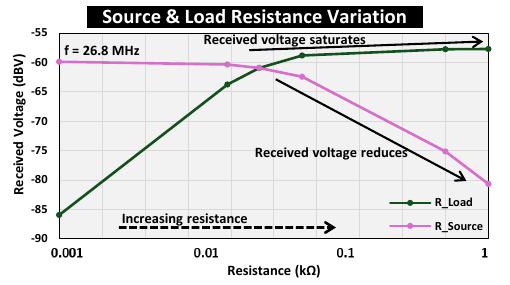}
    \caption{Variation of $R_{Source}$ and $R_{Load}$ at the resonant frequency}
    \label{Impedance}
    \vspace{-0.1in}
\end{figure}

As the source impedance, $R_{Source}$, increases, it will eventually diminish the current passing through the transmitter coil, thereby decreasing the magnetic field strength generated and inducing less current at the receiver. Consequently, the received power is reduced. Therefore, lower source impedance results in higher received voltage. The source impedance plot depicted in Figure \ref{Impedance} illustrates the voltage received at the resonant frequency for various source resistances. To receive a higher voltage, the $R_{Source}$ can be reduced to lower values at the cost of higher power consumption of the transmitter. For a 1V sinusoidal input with $R_{Source}=50ohms$, the power consumed by the transmitter is $10 mW$. 

\subsubsection{Effect of Load Impedance}
Greater load resistance denoted as $R_{Load}$, results in improved received voltage. The elevated resistance at the load directs the majority of the current through the measurement node. As load resistance exceeds 200 ohms, the received voltage saturates, leading to enhanced power transfer. Figure \ref{Impedance} illustrates the received voltage plot at various load resistances.
Communication and powering can be performed by having a dual mode for the $R_{Load}$. The lower the $R_{Load}$, the better it is for powering as discussed in the equation in the above section. Higher $R_{Load}$, enhances voltage mode communication making it a better communication channel. 

\begin{figure}[h]
    \centering
    \includegraphics [width=1\linewidth] {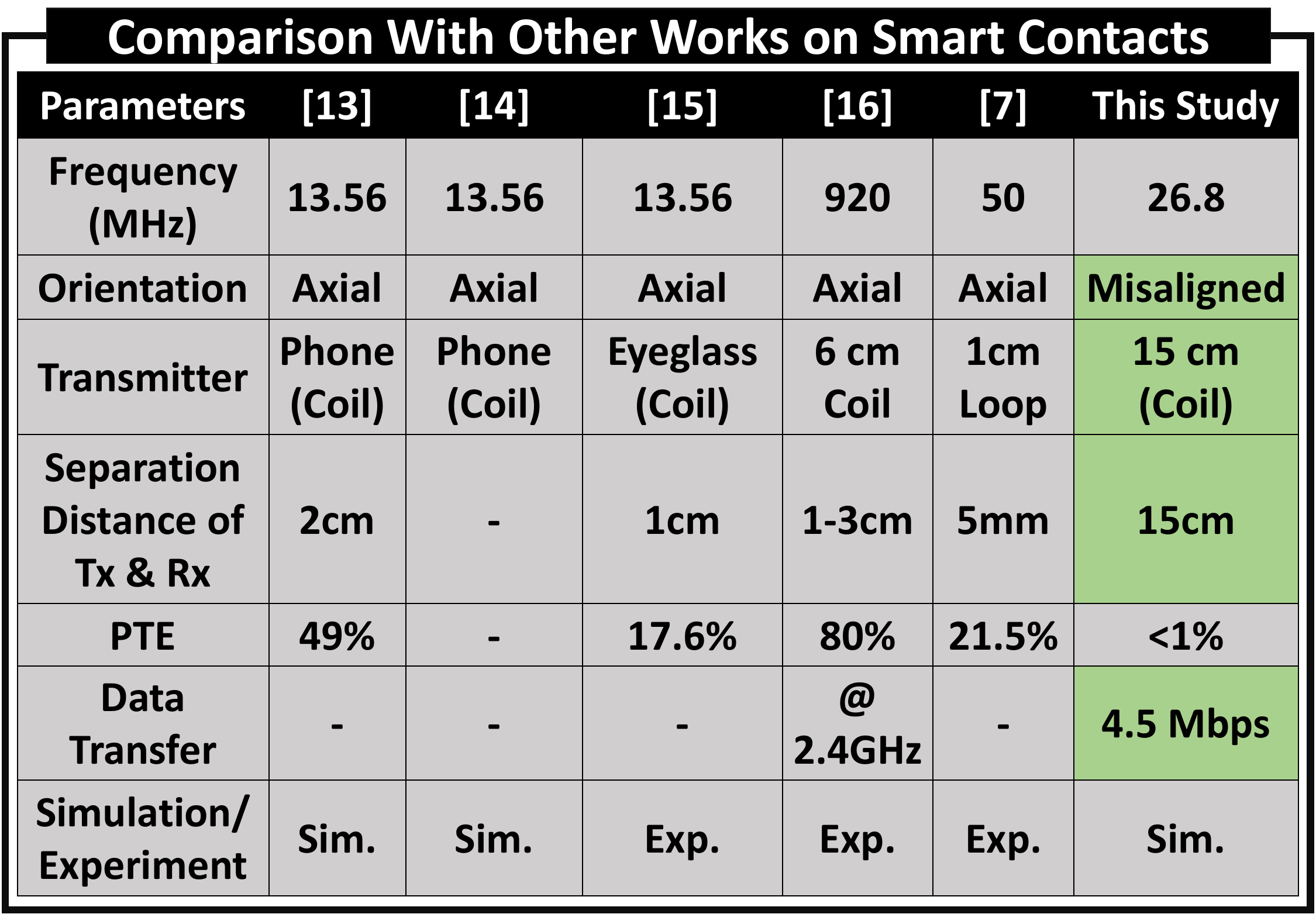}
    \caption{Other techniques that have used inductive coupling for powering and communication of contact lenses. }
    \label{comparison}
    \vspace{-0.2in}
\end{figure}

\section{Discussion}
An inductively coupled two-coiled system is proposed for efficient communication and powering between a necklace and smart contact lens to realize the concept of "Eyes Up, Hands Down" for the vision of having a PC on the eye which not only would help in navigating through documents or grocery lists, but also monitor its user's health using the biomarkers on the eye and deliver drug if needed. The smart contacts can also be used for covert military operations or to detect the movement of the eye while driving. To our knowledge, this is the first approach to show the possibilities of AR/VR on a Smart Contact Lens based on Resonant-MQS field coupling. Despite the misalignments(angular, lateral, and axial) present in the proposed coil systems with a separation distance/channel length of over 15cm, a gain of $-55dB$, and a channel capacity of at least $4.5Mbps$ over a $3dB$ bandwidth of $1MHz$ are achieved. The channel remains unaffected in the presence of body tissues. The power consumption by the transmitter is about $10mW$.  

Several other works on inductively coupled power transfer for contact lenses have been illustrated in Figure \ref{comparison}. The separation distance between the transmitter and receiver coils that have been used for near-field powering and communication does not go over a few centimeters with all of them aligned axially. This work not only shows a high separation distance of $15cm$ but also the robustness against misalignments. There is a lateral shift of over $9cm$ along with that the axis of the receiver coil is perpendicular to that of the transmitter that is inclined at an angle.

\section*{Acknowledgment}
The authors would like to thank the support of their research group members at Purdue University.
\bibliographystyle{IEEEtran}
\bibliography{LaTeX-Template/main}

\end{document}